\documentclass[reprint,prc,showpacs,amsmath,amssymb,superscriptaddress]{revtex4-1}
\usepackage{dcolumn}
\usepackage[dvips]{graphicx}
\usepackage{amsmath}
\usepackage{multirow}
\usepackage{setspace}
\newcommand\T{\rule{0pt}{3.1ex}}
\usepackage{color}
\usepackage{bm}
\usepackage[T1]{fontenc}
\usepackage[latin9]{inputenc}
\usepackage{color}
\usepackage{amssymb}
\usepackage{float}

\usepackage{array}

\begin{document}

\title{Phase space factors and half-life predictions for Majoron emitting $\beta^-\beta^-$ decay} 

\author{J. Kotila}
\email{jenni.kotila@yale.edu}
\affiliation{University of Jyvaskyla, Department of Physics, B.O. Box 35, FI-40014, University of Jyvaskyla, Finland}
\affiliation{Center for Theoretical Physics, Sloane Physics Laboratory,
Yale University, New Haven, Connecticut 06520-8120, USA}
\author{J.\ Barea}
\email{jbarea@udec.cl}
\affiliation{Departamento de F\'{i}sica, Universidad de Concepci\'{o}n,
 Casilla 160-C, Concepci\'{o}n 4070386, Chile}
\author{F. Iachello}
\email{francesco.iachello@yale.edu}
\affiliation{Center for Theoretical Physics, Sloane Physics Laboratory,
Yale University, New Haven, Connecticut 06520-8120, USA}

\begin{abstract}
A complete calculation of phase space factors (PSF) for Majoron emitting $0\nu\beta^-\beta^-$ decay modes is presented. The calculation makes use of exact Dirac wave functions with finite nuclear size and electron screening and includes life-times, single electron spectra, summed electron spectra, and angular electron correlations. Combining these results with recent interacting boson nuclear matrix elements (NME) we make half-life predictions for the the ordinary Majoron decay (spectral index $n$=1). Furthermore, comparing theoretical predictions with the obtained experimental lower bounds for this decay mode we are able to set limits on the effective Majoron-neutrino coupling constant $\langle g_{ee}^M\rangle$.
\end{abstract}

\maketitle
\section{Introduction}

Double-$\beta$ decay is a process in which a nucleus $(A,Z)$ decays to a nucleus $(A,Z\pm2)$ by emitting two electrons or positrons and, usually, other light particles
\begin{equation}
(A,Z)\rightarrow(A,Z\pm 2) + 2e^{\mp} + \text{anything}.
\end{equation}
The mode where two antineutrinos or neutrinos are emitted is predicted by the standard model and has been observed in several nuclei (for a review, see e.g. \cite{bara15}). 
The more exotic mode, neutrinoless double beta decay, is not allowed by the standard model, and once observed would offer new information on many fundamental aspects of elementary particle physics.
As discussed in Ref.~\cite{barea13}, several scenarios of neutrinoless double beta decay have been considered, most notably, light neutrino exchange, heavy neutrino exchange, and Majoron emission.  After the discovery of neutrino oscillations, attention has been focused on the first scenario and the mass mode, where the transition operator is proportional to $\left\langle m_{\nu}\right\rangle/m_e$. 
Even though most current experimental efforts have been focused to the detection of this mode, interest on the  mechanism predicting $0\nu\beta\beta$ decays through the emission of additional bosons called Majorons has also renewed lately.

\begin{figure}[htb!]
\includegraphics[width=1.00\linewidth ]{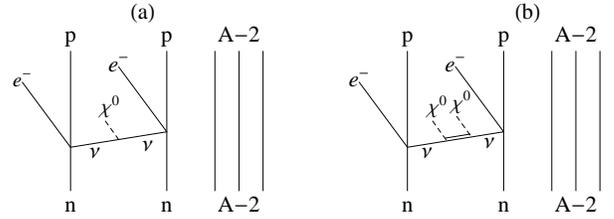} 
\caption{\label{fig1} Neutrinoless double beta decay accompanied by the emission of one or two Majorons.}
\end{figure}

Majorons were introduced years ago \cite{x,xx} as massless Nambu-Goldstone bosons arising from a global $B-L$ (baryon number minus lepton number) symmetry broken spontaneously in the low-energy regime. These bosons couple to the Majorana neutrinos and give rise to neutrinoless double beta decay, accompanied by Majoron emission $0\nu\beta\beta M$ \cite{xxx}, as shown in Fig. \ref{fig1} (a).  
Although these older models are disfavored by precise measurements of the width of the $Z$ boson decay to invisible channels \cite{xv}, several other models of $0\nu\beta\beta M$ decay have been proposed in which one or two Majorons, denoted by $\chi_0$, are emitted, (see Fig. \ref{fig1})
\begin{equation}
(A,Z) \rightarrow(A,Z+2) +2 e^{-} +\chi_0
\end{equation}
or 
\begin{equation}
(A,Z) \rightarrow(A,Z+2) +2 e^{-} +2\chi_0.
\end{equation}
Table \ref{table1} lists some of the models proposed to describe these decays. The different models are distinguished by the nature of the emitted Majoron(s), i.e. is it a Nambu-Goldstone boson or not (NG), the leptonic charge of the emitted Majoron(s) (L), and the spectral index of the model (n), which characterizes the shape of the summed electron spectrum, as described in Sect. \ref{sect2}.
\begin{ruledtabular}
\begin{center}
\begin{table}[h]
\caption{\label{table1}Different Majoron emitting models of $0\nu\beta\beta M$ decay \cite{bam95,hir96,bur95,car93,moh00}. The third, fourth, and fifth  columns indicate whether the Majoron is Nambu-Goldstone boson or not, its leptonic charge $L$, and the model's spectral index $n$.}
\begin{tabular}{lccccc}
Model 	&Decay Mode	&NG boson &L &n & 
\cr \hline 
IB		&$0\nu\beta\beta\chi_0$	&No	&0	&1			\\
IC		&$0\nu\beta\beta\chi_0$	&Yes	&0	&1			\\
ID		&$0\nu\beta\beta\chi_0\chi_0$	&No	&0	&3		\\
IE		&$0\nu\beta\beta\chi_0\chi_0$	&Yes	&0	&3		\\
IIB		&$0\nu\beta\beta\chi_0$	&No	&-2	&1			\\
IIC		&$0\nu\beta\beta\chi_0$	&Yes	&-2	&3			\\
IID		&$0\nu\beta\beta\chi_0\chi_0$	&No	&-1	&3		\\
IIE		&$0\nu\beta\beta\chi_0\chi_0$	&Yes	&-1	&7		\\
IIF		&$0\nu\beta\beta\chi_0$	&Gauge boson	&-2	&3	\\
"Bulk"	&$0\nu\beta\beta\chi_0$	&Bulk field	&0	&2	\\
\end{tabular}
\end{table}
\end{center}
\end{ruledtabular}

In our previous articles we have studied phase space factors (PSF) and prefactors (PF) \cite{kotila12,kotila14,kotila13}, and nuclear matrix elements (NME) \cite{barea,barea12,barea13,barea13b,kotila14, bar15} needed for the theoretical description of $0\nu\beta^-\beta^-$, $2\nu\beta^-\beta^-$,  $0\nu\beta^+\beta^+$, $0\nu\beta^+ EC^+$, $R0\nu ECEC$, $2\nu\beta^+\beta^+$, $2\nu\beta^+EC$, and $2\nu ECEC$ decay mediated in the case of neutrinoless decay by light or heavy neutrino exchange. In this article we continue our systematic evaluation by presenting phase space factors for the different Majoron emitting mechanisms and combining the $n=1$ results with recent interacting boson model (IBM-2) NMEs \cite{bar15} to make half-life predictions for ordinary Majoron decay ($n=1$). Furthermore, we compare our theoretical predictions with the obtained experimental lower bounds for this decay mode to set some limits on the effective Majoron-neutrino coupling constant $\langle g_{ee}^M\rangle$. 

\section{Phase space factors in Majoron emitting double-$\beta$ decay \label {sect2}}
The key ingredients for the evaluation of phase space factors in single- and double-$\beta$ decay are the scattering wave functions and for EC the bound state wave functions. The general theory of relativistic electrons and positrons can be found e.g., in  the book of Rose \cite{rose}. The electron scattering wave functions of interest in $\beta^-\beta^-$ were given in Eq. (8) of~\cite{kotila12}. 

In order to calculate PSFs for Majoron emitting $\beta^-\beta^-$ we use the formulation of Doi, Kotani, and Takasugi \cite{doi}.
The differential rate for the decay is given by \cite{doi,tom91}
\begin{equation}
\label{dw0nu}
dW_{m\chi_0n}=\left(a^{(0)}+a^{(1)}\cos \theta_{12}\right)w_{m\chi_0n}d\epsilon_1d\epsilon_2d(\cos \theta_{12})
\end{equation}
where  $\epsilon_1$ and $\epsilon_2$ are the electron energies,  $\theta_{12}$ the angle between the two emitted electrons, and $w_{m\chi_0n} $ takes different values depending on the number of emitted Majorons $m$ and the spectral index $n$:
\begin{equation}
\begin{split}
w_{1\chi_01}&=\frac{g_A^4 (G\cos\theta_C)^4}{64\pi^7 \hbar}\left(\frac{\hbar c}{2R}\right)^2q(p_1c)(p_2c)\epsilon_1\epsilon_2\\
w_{1\chi_03}&=\frac{g_A^4 (G\cos\theta_C)^4}{64\pi^7 \hbar}q^3(p_1c)(p_2c)\epsilon_1\epsilon_2\\
w_{2\chi_03}&=\frac{g_A^4 (G\cos\theta_C)^4}{3072\pi^9 \hbar}(m_e c^2)^{-2}\left(\frac{\hbar c}{2R}\right)^2q^3(p_1c)(p_2c)\epsilon_1\epsilon_2\\
w_{2\chi_07}&=\frac{g_A^4 (G\cos\theta_C)^4}{53760\pi^9 \hbar}(m_e c^2)^{-6}\left(\frac{\hbar c}{2R}\right)^2q^7(p_1c)(p_2c)\epsilon_1\epsilon_2.\\
\end{split}
\end{equation}
Here $G$ is the Fermi constant, $\theta_C$ is the Cabibbo angle, and the Majoron energy $q$ is determined as $q=Q_{\beta\beta}+2m_ec^2-\epsilon_1-\epsilon_2$, and $R=r_0A^{1/3}$ with $r_0=1.2$ fm, is the nuclear radius.
 The quantities $a^{(0)}$ and $a^{(1)}$ in Eq.~(\ref{dw0nu}) can be written as \cite{tom91}
\begin{equation}
a^{(i)}=f_{11}^{(i)}\left| \langle g_{ee}^M\rangle\right|^{2m} \left| M_{0\nu M}^{(m,n)}\right|^2 \hspace{1cm} i=0,1,
\end{equation}
 where $\left| \langle g_{ee}^M\rangle\right|$ is the effective coupling constant of the Majoron to the neutrino, $m=1,2$ for the emission of one or two Majorons, respectively, and $M_{0\nu M}^{(m,n)}$ is the nuclear matrix element. The functions $f_{11}^{(0)}$, $f_{11}^{(1)}$  are defined as
\begin{equation}
\label{combwave}
\begin{split}
f^{(0)}_{11}&=|f^{-1-1}|^2+|f_{11}|^2+|{f^{-1}}_{ 1}|^2+|{f_1}^{ -1}|^2,\\
f^{(1)}_{11}&=-2\text{Re}[f^{-1-1}f_{11}^*+{f^{-1}}_{ 1}{f_1}^{ -1*}].
\end{split}
\end{equation}
with
\begin{equation}
\begin{split}
f^{-1-1}&=g_{-1}(\epsilon_1)g_{-1}(\epsilon_2),\\
f_{11}&=f_1(\epsilon_1)f_1(\epsilon_2),\\
{f^{-1}}_{ 1}&=g_{-1}(\epsilon_1)f_1(\epsilon_2),\\
{f_{1}}^{ -1}&=f_1(\epsilon_1)g_{-1}(\epsilon_2),
\end{split}
\end{equation}
where $g_{-1}(\epsilon)$ and $f_{1}(\epsilon)$ are obtained from the electron wave functions as explained in Ref. \cite{kotila12}.

All quantities of interest are then given by integration of Eq.~(\ref{dw0nu}). Introducing 
\begin{equation}
\label{gchi}
\begin{split}
G_{m\chi_0n}^{(i)}=&
\frac{2}{g_A^4 \ln2}\int^{Q_{\beta\beta}+m_ec^2}_{m_ec^2}\int^{Q_{\beta\beta}+2m_ec^2-\epsilon_1}_{m_ec^2}f^{(i)}_{11}\\
&\times w_{m\chi_0n}d\epsilon_1d\epsilon_2,
\end{split}
\end{equation}
where the axial vector coupling constant $g_A$ is separated from the phase space factors for conveniency, we can calculate:\\
(i) The half-life
\begin{equation}
\left[ \tau^{0\nu}_{1/2} \right]^{-1}=g_A^4G_{m\chi_0n}^{(0)} \left|\left\langle g_{\chi_{ee}^M}\right\rangle\right|^{2m} \left| M_{0\nu M}^{(m,n)}\right|^2,
\end{equation}
(ii) the single electron spectrum
\begin{equation}
\frac{dW_{m\chi_0n}}{d\epsilon_1}={\cal N}_{0\nu M}^{(m,n)} \frac{dG_{m\chi_0n}^{(0)}}{d\epsilon_1}
\end{equation}
where ${\cal N}_{0\nu M}^{(n,m)}=g_A^4ln2\left|\langle g_{\chi_{ee}^M}\rangle\right|^{2m}\left| M_{0\nu M}^{m,n}\right|^2$.\\
(iii) The summed electron spectrum, which shape makes the different Majoron emitting modes experimentally recognizable
\begin{equation}
\frac{dW_{m\chi_0n}}{d(\epsilon_1+\epsilon_2)}={\cal N}_{0\nu M}^{(n,m)} \frac{dG_{m\chi_0n}^{(0)}}{d(\epsilon_1+\epsilon_2)},
\end{equation}
(iv) and the angular correlation between the two electrons
\begin{equation}
\alpha(\epsilon_1)=
\frac{dG_{m\chi_0n}^{(1)}/d\epsilon_1}{dG_{m\chi_0n}^{(0)}/d\epsilon_1}.
\end{equation}
\begin{ruledtabular}
\begin{center}
\begin{table}[h]
\caption{\label{0nuG}Phase space factors $G_{m\chi_0n}^{(0)}$ obtained using screened exact finite size Coulomb wave functions.}
\begin{tabular}{lcccc}
 	&\multicolumn{4}{c}{$G_{m\chi_0n}^{(0)}$$(10^{-18}$ yr$^{-1})$}\\\cline{2-5}

Nucleus		&m=1,n=1 	&m=1,n=3	&m=2,n=3	&m=2,n=7\\
\cr \hline 
$^{48}$Ca		&1540	&17.1	&73.6	&690 \\
$^{76}$Ge	&44.2	&0.073	&0.22	&0.420\\
$^{82}$Se		&361		&1.22	&3.54	&26.9 \\
$^{96}$Zr		&905		&4.21	&11.0	&128.\\
$^{100}$Mo	&598		&2.42	&6.15	&50.8 \\
$^{110}$Pd	&94.1	&0.205	&0.487	&0.946 \\
$^{116}$Cd	&569		&2.28	&5.23	&33.9 \\
$^{124}$Sn	&209		&0.653	&1.45	&4.45 \\
$^{128}$Te	&3.06	&0.001	&0.003	&0.0003 \\
$^{130}$Te	&413		&1.51	&3.21	&14.4 \\
$^{134}$Xe	&2.92	&0.002	&0.003	&0.0002	\\
$^{136}$Xe	&409		&1.47	&3.05	&12.5 \\
$^{148}$Nd	&197		&0.505	&0.986	&1.72\\
$^{150}$Nd	&3100	&21.1	&40.8	&538\\
$^{154}$Sm	&28.2	&0.034	&0.064	&0.021 \\
$^{160}$Gd	&1590	&0.361	&0.672	&0.899 \\
$^{198}$Pt	&60.7	&0.068	&0.110	&0.021 \\
$^{232}$Th	&82.4	&0.073	&0.105	&0.009 \\
$^{238}$U		&337		&0.532	&0.756	&0.213 \\
\end{tabular}
\end{table}
\end{center}
\end{ruledtabular}

\begin{figure}[htb!]
\includegraphics[width=1.00\linewidth ]{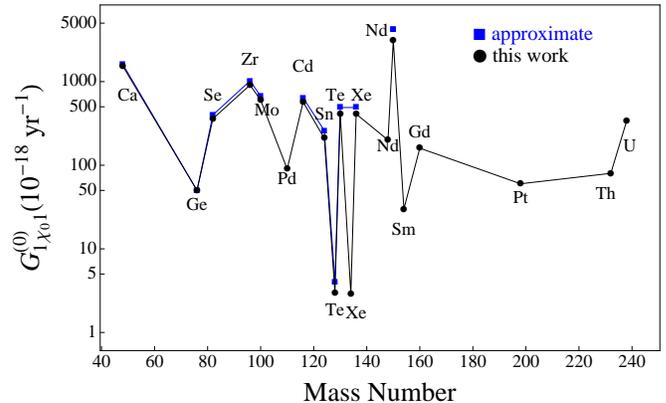} 
\caption{\label{Gcomp1}(Color online) Phase space factors $G_{1\chi_01}^{(0)}$ in units $(10^{-18}$ yr$^{-1})$. The label "approximate" refers to the results obtained by the use of approximate electron wave functions \cite{doi, tom91, suhonen}.  The figure is in semilogarithmic scale.}
\end{figure}

\begin{figure}[htb!]
\includegraphics[width=1.00\linewidth ]{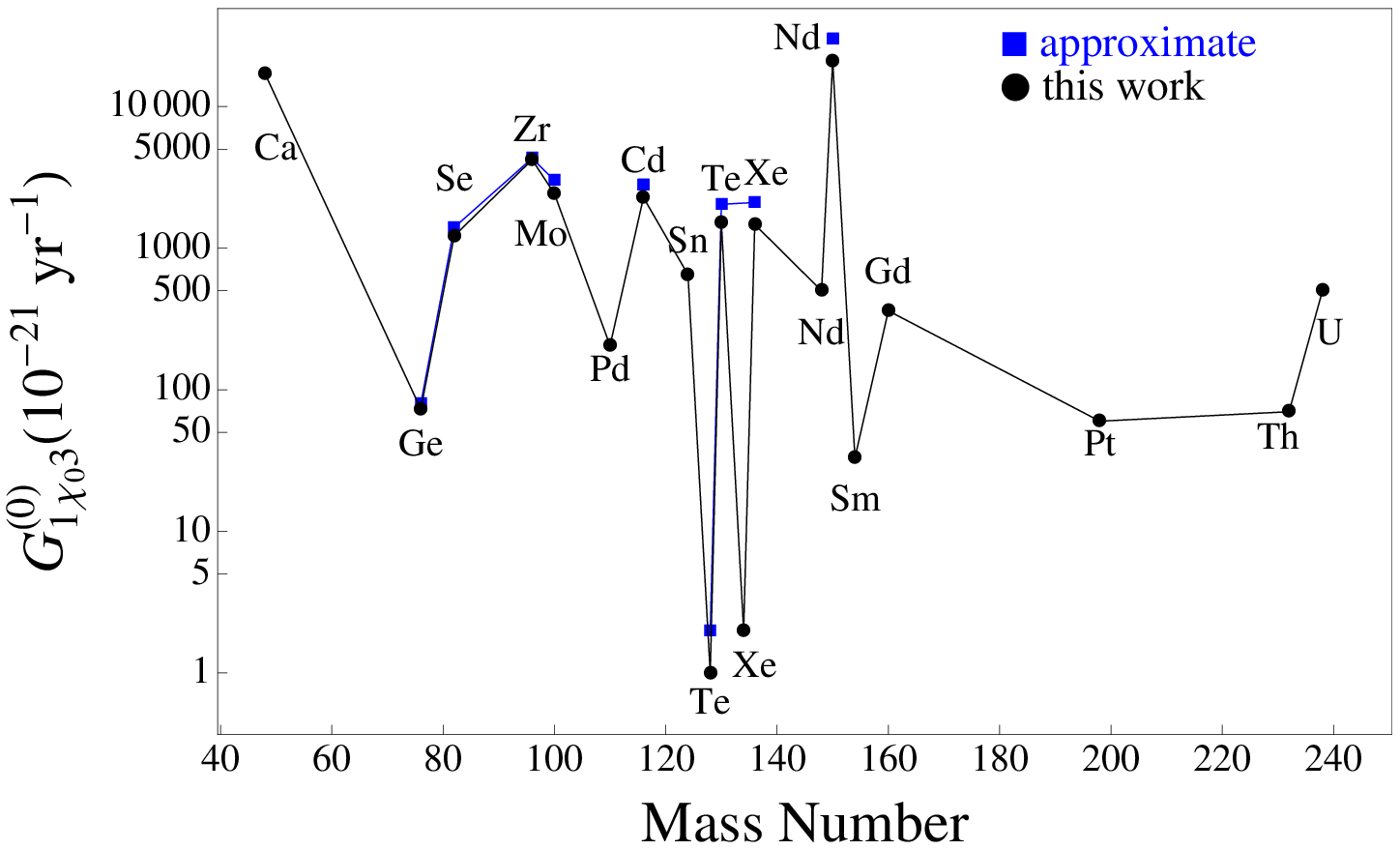} 
\caption{\label{Gcomp2}(Color online) Phase space factors $G_{1\chi_03}^{(0)}$ in units $(10^{-21}$ yr$^{-1})$. The label "approximate" refers to the results obtained by the use of approximate electron wave functions \cite{gun96,arn00}.  The figure is in semilogarithmic scale.}
\end{figure}

\begin{figure}[htb!]
\includegraphics[width=1.00\linewidth ]{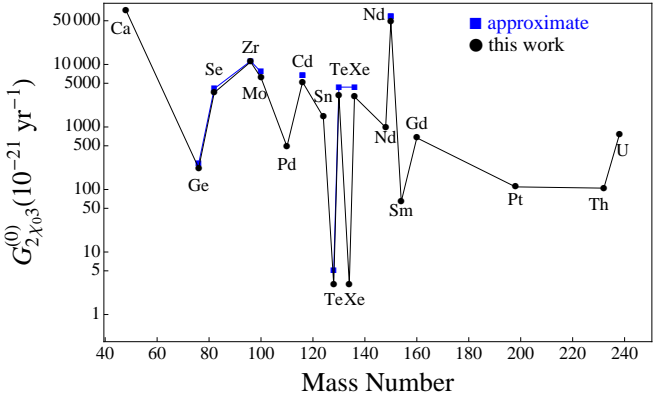} 
\caption{\label{Gcomp3}(Color online) Same as Fig. \ref{Gcomp2} for  $G_{2\chi_03}^{(0)}$. }
\end{figure}

\begin{figure}[htb!]
\includegraphics[width=1.00\linewidth ]{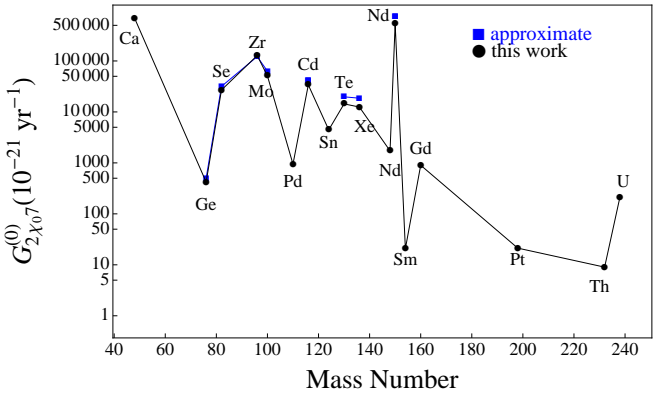} 
\caption{\label{Gcomp4}(Color online) Same as Fig. \ref{Gcomp2}  for $G_{2\chi_07}^{(0)}$. Due to a very small values, $^{128}$Te and  $^{134}$Xe are excluded from this figure.}
\end{figure}

We have done a calculation of $G_{m\chi_0n}^{(0)}$ and $G_{m\chi_0n}^{(1)}$ in the list of nuclei shown in Table~\ref{0nuG}.  We also plot our results in Figs.~\ref{Gcomp1}-\ref{Gcomp4}, where  they are compared with previous calculations \cite{doi,tom91, suhonen, gun96,arn00}. For the comparison the values of \cite{doi, suhonen} have been multiplied by a missing factor of two and divided by $g_A^4$, and values of \cite{tom91} have been divided with factor $g_A^44R^2$.  The factor of two is the correct choice, as was acknowledged in \cite{doi88}.   This factor of two is also included in Table II of Ref. \cite{alb14}, where the decay of $^{136}$Xe have been studied. In their calculation for the phase space factors they use Fermi functions F(Z,E) that fully include nuclear finite size and electron screening and are evaluated at the nuclear radius. Their result with different $m=1,2$ and $n=1,3,7$ are within 6\% of the ones reported here. 


\begin{figure*}[ctb!]
\begin{center}
\begin{tabular}{c}
\includegraphics[width=1.00\linewidth]{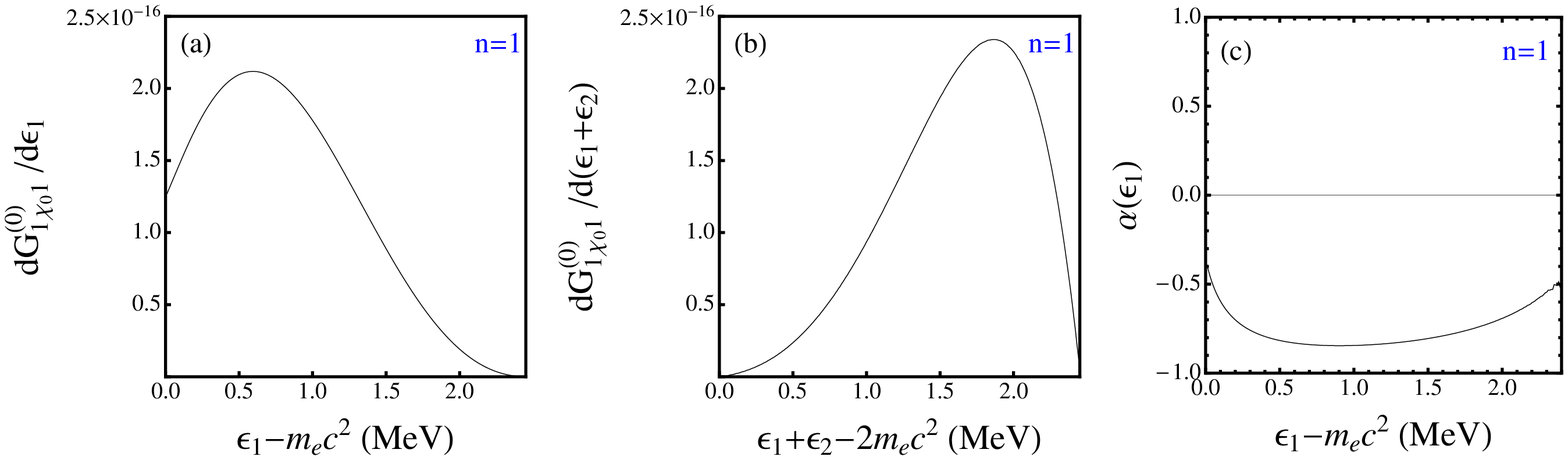} \\
\end{tabular}
\begin{tabular}{c}
\includegraphics[width=1.00\linewidth]{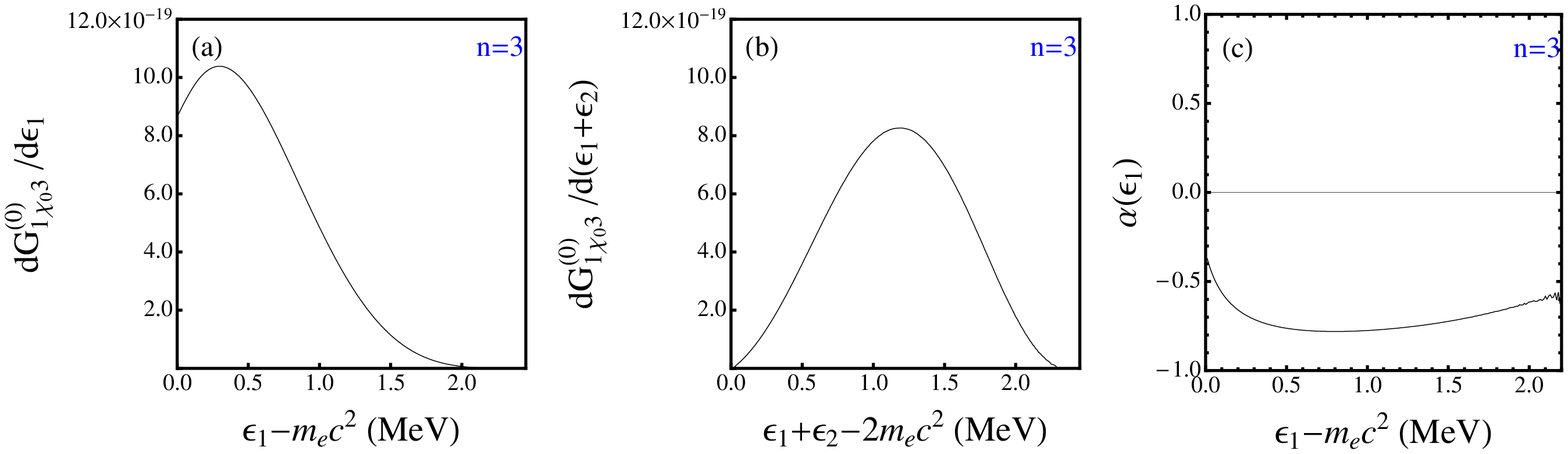} \\
\end{tabular}
\begin{tabular}{c}
\includegraphics[width=1.00\linewidth]{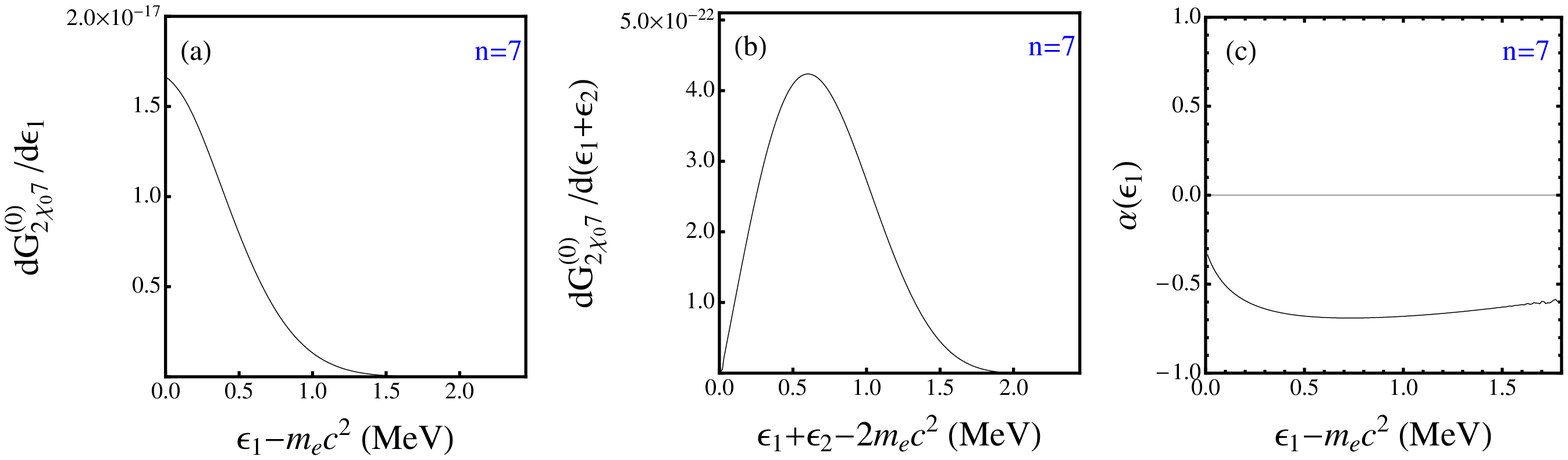} \\
\end{tabular}
\caption{(Color online)\label{xe}Single electron spectra (panel a), summed electron spectra (panel b), and angular correlations between the two outgoing electrons (panel c) for the  $^{136}$Xe $\rightarrow ^{136}$Ba $0\nu\beta\beta M$-decay with different spectral indices $n=1,3,7$. The scale of the panels (a) and (b) should be multiplied by ${\cal N}_{0\nu M}^{(n,m)}$ for a realistic estimate.}
\label{xe}
\end{center}
\end{figure*}

We also have available upon request the single and summed electron electron spectra and angular correlation for all nuclei in Table~\ref{0nuG}. An example, $^{136}$Xe decay, is shown in Fig.~\ref{xe}. The shape of the spectra is determined by the spectral index $n$, i.e. for the case $n=3$ it is the same for both the emission of one or two Majorons, but the overall scale is different ($\chi_0$ result should be multiplied by $(\hbar c)^2(2Rm_ec^2)^{-2}(48\pi^2)^{-1}$ to obtain $2\chi_0$ result). Also, since the angular correlation is obtained by dividing $dG_{m\chi_0n}^{(1)}/d\epsilon_1$ by $dG_{m\chi_0n}^{(0)}/d\epsilon_1$ the calculation becomes unstable for the higher $n$ near the endpoint energy and that region is thus excluded from Fig. \ref{xe}. The distinction between different values of $n$ is most prominent in summed electron spectra, where as $n$ increases the peak of the spectrum shifts from near the maximum kinetic energy to near the minimum kinetic energy. The difference between different values of $n$ is also shown in single electron spectra but not as strongly as in the case of summed electron spectra.

\section{Expected half-lives and limits on coupling constant: Ordinary Majoron emitting $\beta^-\beta^-$ decay}

\begin{figure}[h!]
\begin{center}
\includegraphics[width=1.\linewidth]{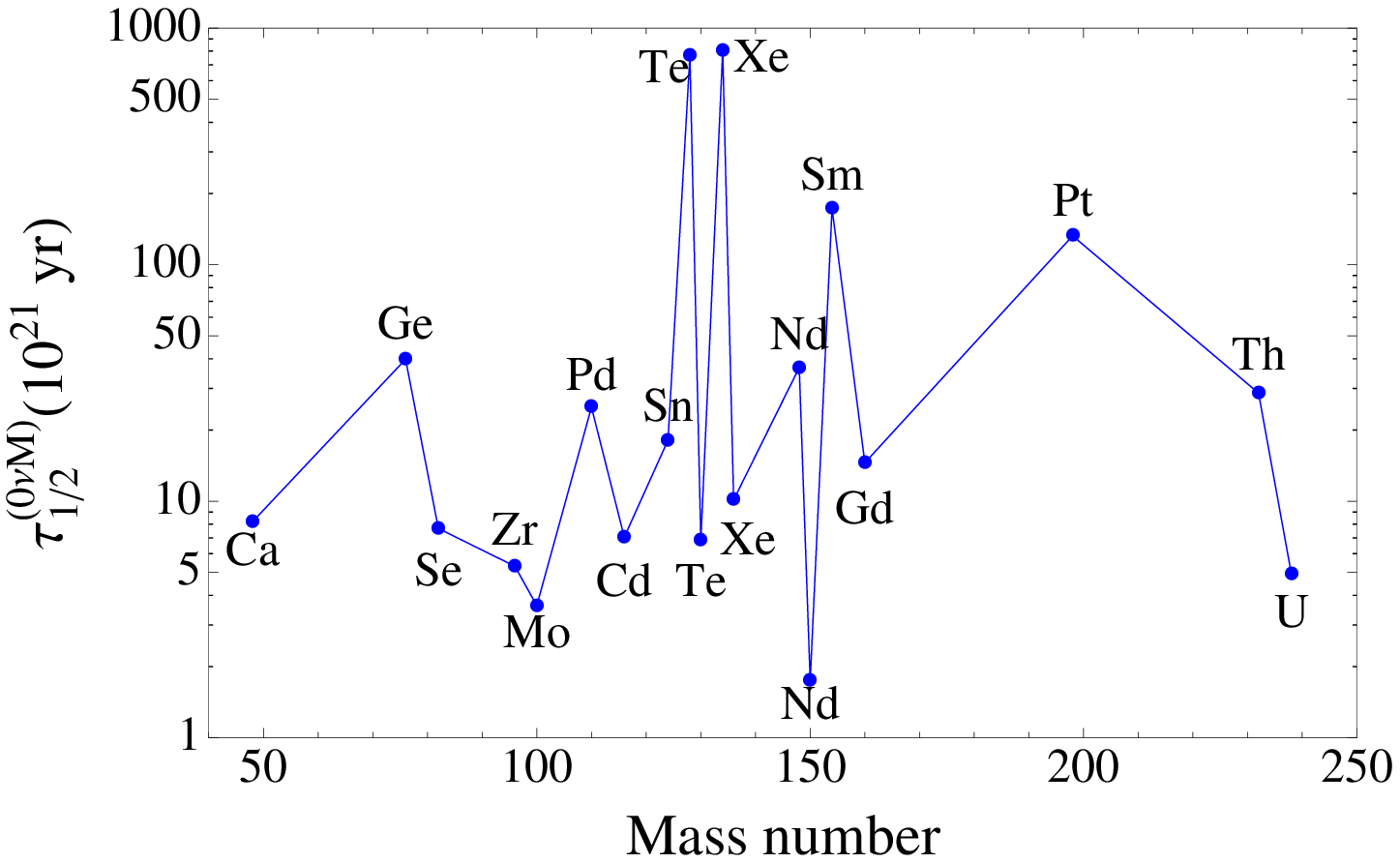} 
\end{center}
\caption{\label{fig6}(Color online) Expected half-lives for Majoron decay models where $n=1$ (IB,IC,IIB)  with $\left\langle g_{ee}^M\right\rangle=10^{-4}$, $g_{A}=1.269$ and IBM-2 isospin restored nuclear matrix elements. The figure is in semilogarithmic scale.}
\end{figure}

In the case of ordinary Majoron emitting double-$\beta$ decay, $n=1$, the nuclear matrix elements have the same form as in the $0\nu\beta\beta$ mediated by light neutrino exchange
\begin{equation}
\label{sum}
M_{0\nu M}^{(1,1)}=
g_A^2\left(M_{GT}^{(0\nu)} -\left(\frac{g_{V}}{g_{A}}\right)^2 M_{F}^{(0\nu)}+M_{T}^{(0\nu)}\right).
\end{equation}
The calculation  of  phase space factors can now be combined with updated nuclear matrix elements in IBM-2 \cite{bar15} to produce predictions for half-lives for ordinary Majoron decay ($n=1$) in Table~\ref{table3} (left) and Fig.~\ref{fig6}. 
Judging by the predicted half-lives, the most prominent candidates are $^{150}$Nd and $^{100}$Mo. Furthermore, we can compare our predictions to half-life limits coming from experiments to set some limits on the effective Majoron-neutrino coupling constant. The obtained limits are shown on Table~\ref{table3} (right). The most stringent limits for ordinary Majoron decay ($n=1$) at the moment are for $^{136}$Xe coming from KamLAND-Zen \cite{gan12} and EXO-200 \cite{alb14} experiments reaching the order of magnitude of $10^{-6}$.

\begin{ruledtabular}
\begin{table}[h!]
\caption{\label{table3}Left: Calculated half-lives for Majoron decay models where $n=1$ (IB, IC, IIB)  with $\left\langle g_{ee}^M\right\rangle=10^{-4}$, $g_{A}=1.269$ and recent IBM-2  nuclear matrix elements \cite{bar15}. Right: Upper limit on Majoron coupling constant  $\left\langle g_{ee}^M\right\rangle$ from current experimental limits.}
\begin{tabular}{lc|cc}
Decay  &  \ensuremath{\tau_{1/2}^{0\nu M}}(\ensuremath{10^{21}}yr) &  \ensuremath{\tau_{1/2, exp}^{0\nu M}}(yr) &$\left< g_{ee}^M\right>$ (eV)\\
 \hline
 \T
$^{48}$Ca$\rightarrow ^{48}$Ti		&8.19 	&$>7.2\times 10^{20}$\footnotemark[1] &$<3.4\times 10^{-4}$\\
$^{76}$Ge$\rightarrow ^{76}$Se 	&39.8 	&$>6.4\times 10^{22}$\footnotemark[2] &$<7.9\times 10^{-5}$\\
$^{82}$Se$\rightarrow ^{82}$Kr	 	&7.68 	&$>1.5\times 10^{22}$\footnotemark[3] &$<7.2\times 10^{-5}$\\
$^{96}$Zr$\rightarrow ^{96}$Mo	&5.32 	&$>1.9\times 10^{21}$\footnotemark[3] &$<1.7\times 10^{-4}$\\
$^{100}$Mo$\rightarrow ^{100}$Ru 	&3.62 	&$>3.9\times 10^{22}$\footnotemark[4] &$<3.0\times 10^{-5}$\\
$^{110}$Pd$\rightarrow ^{110}$Cd 	&25.0 	& &\\
$^{116}$Cd$\rightarrow ^{116}$Sn  	&7.06 	&$>8\times 10^{21}$\footnotemark[5] &$9.4\times 10^{-5}$\\
$^{124}$Sn$\rightarrow ^{124}$Te 	&18.1 	& &\\
$^{128}$Te$\rightarrow ^{128}$Xe 	&765 	&$>2\times 10^{24}$\footnotemark[6] &$<6.2\times 10^{-5}$\\
$^{130}$Te$\rightarrow ^{130}$Xe 	&6.82 	&$>1.6\times 10^{22}$\footnotemark[7] &$<6.5\times 10^{-5}$\\
$^{134}$Xe$\rightarrow ^{134}$Ba 	&805 & &\\
$^{136}$Xe$\rightarrow ^{136}$Ba 	&10.1 	&$>2.6\times 10^{24}$\footnotemark[8] &$<6.2 \times 10^{-6}$\\
							&	 	&$>1.2\times 10^{24}$\footnotemark[9] &$<9.2\times 10^{-6}$\\

$^{148}$Nd$\rightarrow ^{148}$Sm 	&36.8 & &\\
$^{150}$Nd$\rightarrow ^{150}$Sm 	&1.74 &	$>1.5\times 10^{21}$\footnotemark[3] &$<1.1\times 10^{-4}$\\
$^{154}$Sm$\rightarrow ^{154}$Gd 	&173 & &\\
$^{160}$Gd$\rightarrow ^{160}$Dy 	&14.5 & &\\
$^{198}$Pt$\rightarrow ^{198}$Hg 	&132 & &\\
$^{232}$Th$\rightarrow ^{232}$U 	&28.7\\
$^{238}$U$\rightarrow ^{238}$Pu 	&4.94\\
\end{tabular}
\footnotetext[1]{Ref.~\cite{bar89}.}
\footnotetext[2]{Ref.~\cite{kla01}.}
\footnotetext[3]{Ref.~\cite{bar11}.}
\footnotetext[4]{Ref.~\cite{arn14}.}
\footnotetext[5]{Ref.~\cite{dan03}.}
\footnotetext[6]{Ref.~\cite{man91}, geochemical.}
\footnotetext[7]{Ref.~\cite{arn11}.}
\footnotetext[8]{Ref.~\cite{gan12}.}
\footnotetext[9]{Ref.~\cite{alb14}.}
\end{table}
\end{ruledtabular}

\section{Conclusions}
In this article, we have reported a complete  calculation of phase space factors for $0\nu\beta^-\beta^-$ decay proceeding through emission of one or two Majorons. The reported results include half-lives, single electron spectra, summed electron spectra, and electron angular correlations, to be used in connection with the calculation of nuclear matrix elements. Furthermore, we have combined our results with recent IBM-2 nuclear matrix elements to produce predictions of half-lives in the case of ordinary Majoron decay, spectral index $n=1$. Comparing these  predictions with experimental lower bounds, we have set some limits on the effective Majoron-neutrino coupling constant $\langle g_{ee}^M\rangle$. At the moment the best limits are coming  from $^{136}$Xe  experiments reaching the order of magnitude of $10^{-6}$. Also, the results in Table III are for $g_A=1.269$. If $g_A$ is renormalized to $g_{A,eff}$, $\tau_{1/2}^{0\nu M}$ should be multiplied by $(1.269/g_{A.eff})^4$ and limits on $\langle g_{ee}^M\rangle$ by $(1.269/g_{A.eff})^2$.

\acknowledgments
This work was supported in part by US Department of Energy (Grant No. DE-FG-02-91ER-40608), Chilean
Ministry of Education (Fondecyt Grant No. 1150564), Academy of Finland (Project 266437), and by the facilities and staff of the Yale University Faculty of Arts and Sciences High Performance Computing Center.


\begin{thebibliography}{99}
\bibitem{bara15} A. S. Barabash, Nucl. Phys. A \textbf{935}, 52 (2015).
\bibitem{barea13} J.\ Barea, J. Kotila and F.\ Iachello, Phys. Rev. C \textbf{87},
014315 (2013).

\bibitem{x} Y. Chikashige, R. N. Mohapatra, and R. D. Peccei, Phys. Rev Lett. \textbf{45}, 1926 (1980).
\bibitem{xx}  G. B. Gelmini and M. Roncadelli, Phys. Lett. B \textbf{99}, 411 (1981).
\bibitem{xxx}  H. M. Georgi, S. L. Glashow, and S. Nussinov, Nucl. Phys. B \textbf{193}, 297 (1981).
\bibitem{xv}  The ALEPH Collaboration, The DELPHI Collaboration, The L3 Collaboration, The OPAL Collaboration, The SLD Collaboration, The LEP Electroweak Working Group, The SLD Electroweak and Heavy Flavour Groups, Phys. Rep. \textbf{427}, 257 (2006).


\bibitem{bam95}P. Bamert, C. Burgess, and R. Mohapatra, Nucl. Phys. B \textbf{449}, 25 (1995).
\bibitem{car93}C. D. Carone, Phys. Lett. B \textbf{308}, 85 (1993).
\bibitem{hir96} M. Hirsch, H. V. Klapdor-Kleingrothaus, S. G. Kovalenko, and H. Päs, Phys. Lett. B \textbf{372}, 8 (1996).
\bibitem{bur95}C. Burgess and J. Cline, in Proceedings of the First International Conference on Nonaccelerator Physics, Bangalore, India, 1994, edited by R. Cowsik (Wolrd Scientific, Singapore, 1995).
\bibitem{moh00}R. Mohapatra, A. Pe\'rez-Lorenzana, and C. D. S. Pires, Phys. Lett. B \textbf{491}, 143 (2000).
\bibitem{kotila12}J. Kotila and F. Iachello, Phys. Rev. C \textbf{85}, 034316 (2012).
\bibitem{kotila14} J. Kotila, J. Barea, and F.\ Iachello, Phys. Rev. C  \textbf{89}, 064319 (2014).
\bibitem{kotila13} J. Kotila and F.\ Iachello, Phys. Rev. C \textbf{87},
024313 (2013).
\bibitem{barea} J.\ Barea and F.\ Iachello, Phys. Rev. C \textbf{79},
044301 (2009).
\bibitem{barea12} J.\ Barea, J. Kotila, and F.\ Iachello, Phys. Rev. Lett. \textbf{109},
042501 (2012).
\bibitem{barea13b} J.\ Barea, J. Kotila and F.\ Iachello, Phys. Rev. C \textbf{87},
057301 (2013).
\bibitem{bar15}J.\ Barea, J.\ Kotila and F.\ Iachello, Phys. Rev. C \textbf{91}, 034304 (2015).
\bibitem{rose} M.E. Rose, \textit{Relativistic Electron Theory} (Wiley, New York, 1961).
\bibitem{doi} M. Doi, T. Kotani, and E. Takasugi, Prog. Theor. Phys. suppl. \textbf{83}, 1 (1985).
\bibitem{tom91} T. Tomoda, Rep. Prog. Phys. \textbf{54}, 53 (1991).
\bibitem{suhonen} J.  Suhonen and O. Civitarese, Phys. Rep. \textbf{300}, 123 (1998).
\bibitem{gun96} M. Gunther \textit{et al.}, Phys. Rev. D \textbf{54}, 3641 (1996), J. Helmig \textit{et al.}, in Proc. Int. Workshop Double Beta Decay and Related Topics, World Scientific, Singapore, 1996.
\bibitem{arn00} R. Arnold \textit{et al.} (NEMO-2 Collaboration), Nucl. Phys. A,  \textbf{678}, 341 (2000).
\bibitem{doi88} M. Doi, T. Kotani, and E. Takasugi, Phys. Rev. D \textbf{37}, 2575 (1988).
\bibitem{alb14} J. B. Albert \textit{et al.} (EXO-200 Collaboration), Phys. Rev. D \textbf{90}, 092004 (2014).
\bibitem{bar89} A. S. Barabash, Phys. Lett. B \textbf{216}, 257 (1989).
\bibitem{kla01} H. V. Klapdor-Kleingrothaus \textit{et al.} , Eur. Phys. J. A  \textbf{12}, 147 (2001).
\bibitem{bar11} A. S. Barabash and V. B. Brudanin (NEMO Collaboration), Phys. At. Nucl. \textbf{74}, 312 (2011).
\bibitem{arn14} R. Arnold \textit{et al.} (NEMO-3 collabaration), Phys. Rev. D  \textbf{89}, 111101 (2014).


\bibitem{dan03} F. A. Danevich \textit{et al.}, Phys. Rev. C \textbf{68}, 035501 (2003).
\bibitem{man91} O. K. Manuel, J. Phys. G  \textbf{17}, 221 (1991).

\bibitem{arn11} R. Arnold \textit{et al.} (NEMO-3 Collaboration), Phys. Rev. Lett.  \textbf{107}, 062504 (2011).
\bibitem{gan12} A. Gando \textit{et al.} (KamLAND-Zen Collaboration), Phys. Rev. C \textbf{86}, 021601(R) (2012).

\end{thebibliography}
\end{document}